\documentclass[prb,reprint,superscriptaddress,amsmath,amssymb,aps,]{revtex4-2}
\usepackage[utf8]{inputenc}
\usepackage[T1]{fontenc}

\usepackage{amsmath}
\usepackage{amssymb}
\usepackage{graphicx}
\usepackage{dcolumn}
\usepackage{verbatim}
\usepackage{color}
\newcommand{\eis}{Eu$_5$In$_2$Sb$_6$}
\DeclareUnicodeCharacter{2212}{-}

\begin{document}
%\preprint{APS/123-QED}

\title{Magnetic and electronic properties of Eu$_5$In$_2$Sb$_6$}

\author{M. Victoria Ale Crivillero}
 \affiliation{Max Planck Institute for Chemical Physics of Solids, Nöthnitzer Strasse 40, 01187 Dresden, Germany}
\author{Sahana Rößler}
 \affiliation{Max Planck Institute for Chemical Physics of Solids, Nöthnitzer Strasse 40, 01187 Dresden, Germany}
\author{S. Granovsky}
 \affiliation{Institut f\"ur Festk\"orper- und Materialphysik, Technische
 Universit\"at Dresden, D-01062 Dresden, Germany}
\author{M. Doerr}
 \affiliation{Institut f\"ur Festk\"orper- und Materialphysik, Technische
 Universit\"at Dresden, D-01062 Dresden, Germany}
 \author{M. S. Cook}
 \affiliation{Los Alamos National Laboratory, Los Alamos, New Mexico 87545, USA}
 \author{Priscila F. S. Rosa}
 \affiliation{Los Alamos National Laboratory, Los Alamos, New Mexico 87545, USA}
\author{J. Müller}
 \affiliation{Institute of Physics, Goethe-University Frankfurt, 60438
 Frankfurt(M), Germany}
\author{S. Wirth}
 \email{Steffen.Wirth@cpfs.mpg.de}
 \affiliation{Max Planck Institute for Chemical Physics of Solids, Nöthnitzer Strasse 40, 01187 Dresden, Germany}
\date{\today}
%\keywords{Keyword1, Keyword2, Keyword3}

\begin{abstract}
The intermetallic compound \eis, an antiferromagnetic material with
nonsymmorphic crystalline structure, is investigated by magnetic, electronic
transport and specific heat measurements. Being a Zintl phase, insulating
behavior is expected. Our thermodynamic and magnetotransport measurements
along different crystallographic directions strongly indicate polaron
formation well above the magnetic ordering temperatures. Pronounced
anisotropies of the magnetic and transport properties even above the magnetic
ordering temperature are observed despite the Eu$^{2+}$ configuration which
testify to complex and competing magnetic interactions between these ions
and give rise to intricate phase diagrams discussed in detail. Our results
provide a comprehensive framework for further detailed study of this
multifaceted compound with possible nontrivial topology.
\end{abstract}
\maketitle

\section*{Introduction}
Materials in which the electronic and magnetic properties are strongly coupled
hold great promise for applications in spin electronics \cite{zie01}. From a
fundamental point of view, the underlying entanglement of magnetic,
electronic and structural degrees of freedom in correlated electron materials
can give rise to unexpected and often spectacular physical phenomena, e.g.\
high-temperature superconductivity in cuprates \cite{kei15} and colossal
magnetoresistance (CMR) in magnetic semiconductors and manganites \cite{coe99,
kam02}. A characteristic of these coupled degrees of freedom is the appearance
of diverse electronic phases, along with phase separation and pattern formation
\cite{kiv03,cox07,tok17,mat16}. In this respect, Eu compounds are of particular
interest because of their often strong exchange interaction between the charge
carriers and the localized spin of Eu$^{2+}$ ions \cite{mol67,kas68,oli72},
which can give rise to a localization of the charge carriers (of low density)
around the Eu$^{2+}$ spins---so called magnetic polarons. In addition, the
ground state of Eu$^{2+}$ is an isotropic $^8 S_{7/2}$ configuration, limiting
crystalline electric field contributions to anisotropy to higher order. In
general, Eu compounds can be viewed as model systems for CMR effects and
polaron formation \cite{urb04,poh18,sho19}.

Considering Eu-based compounds, research concentrated on EuO, Eu chalcogenides
and EuB$_6$, mostly because of their relatively simple crystallographic
structures \cite{oli72,tor72,sue98}. Recently, the topological properties of
materials and their relation to crystal symmetry has been highlighted
\cite{zha19a,tan19}. Specifically, materials crystallizing in nonsymmorphic
space groups are prone to topological states \cite{par13,wie18}. Interestingly,
space group $Pbam$ (No. 55) fulfills these requirements \cite{wie18} including
the Zintl phase Ba$_5$In$_2$Sb$_6$ \cite{cor88}. This drew attention to the
Eu-based counterpart \eis\ \cite{ros20} which was known as a narrow gap
semiconductor \cite{par02}. Indeed, \eis\ exhibits an extraordinarily large
CMR effect and signatures of polaron formation, with the possibility of axion
insulating states within the antiferromagnetic regime \cite{ros20}. In
addition, band structure calculations for \eis\ in its paramagnetic state
predicted nontrivial surface states \cite{yxu22} (similar to the isostructural
nonmagnetic compound Ba$_5$In$_2$Sb$_6$ \cite{wie18}), which have not been
observed \cite{ale22}.

However, the complex structure of \eis\ allows for three crystallographically
different Eu sites \cite{par02}, and the Eu sublattices may magnetically order
independently \cite{ale22}. In this case, the Dzyaloshinskii–Moriya interaction
needs to be taken into consideration. Independent of the crystallographic site,
Eu is divalent \cite{rad20} and hence, fulfills the Zintl rule. It is
noteworthy that, even though the Eu$^{2+}$ ions with orbital angular momentum
$L = 0$ are isotropic, there appear to be multiple, anisotropic exchange
interactions \cite{ros20,ale22}. All this results in a complex magnetic
structure which is not yet resolved experimentally.

Here we report on a comprehensive study of \eis\ by magnetic, electronic
transport and specific heat measurements in an effort to establish the magnetic
field--temperature ($H$--$T$) phase diagrams with $H$ applied along different
crystallographic directions. The former two properties are highly anisotropic
despite the Eu$^{2+}$ state indicating complex magnetic interactions. The
temperature evolution of the studied properties establish \eis\ as a rare
example of a material exhibiting polaron formation in an antiferromagnet,
\begin{figure*}[t]
\centering
\includegraphics[width=0.76\textwidth]{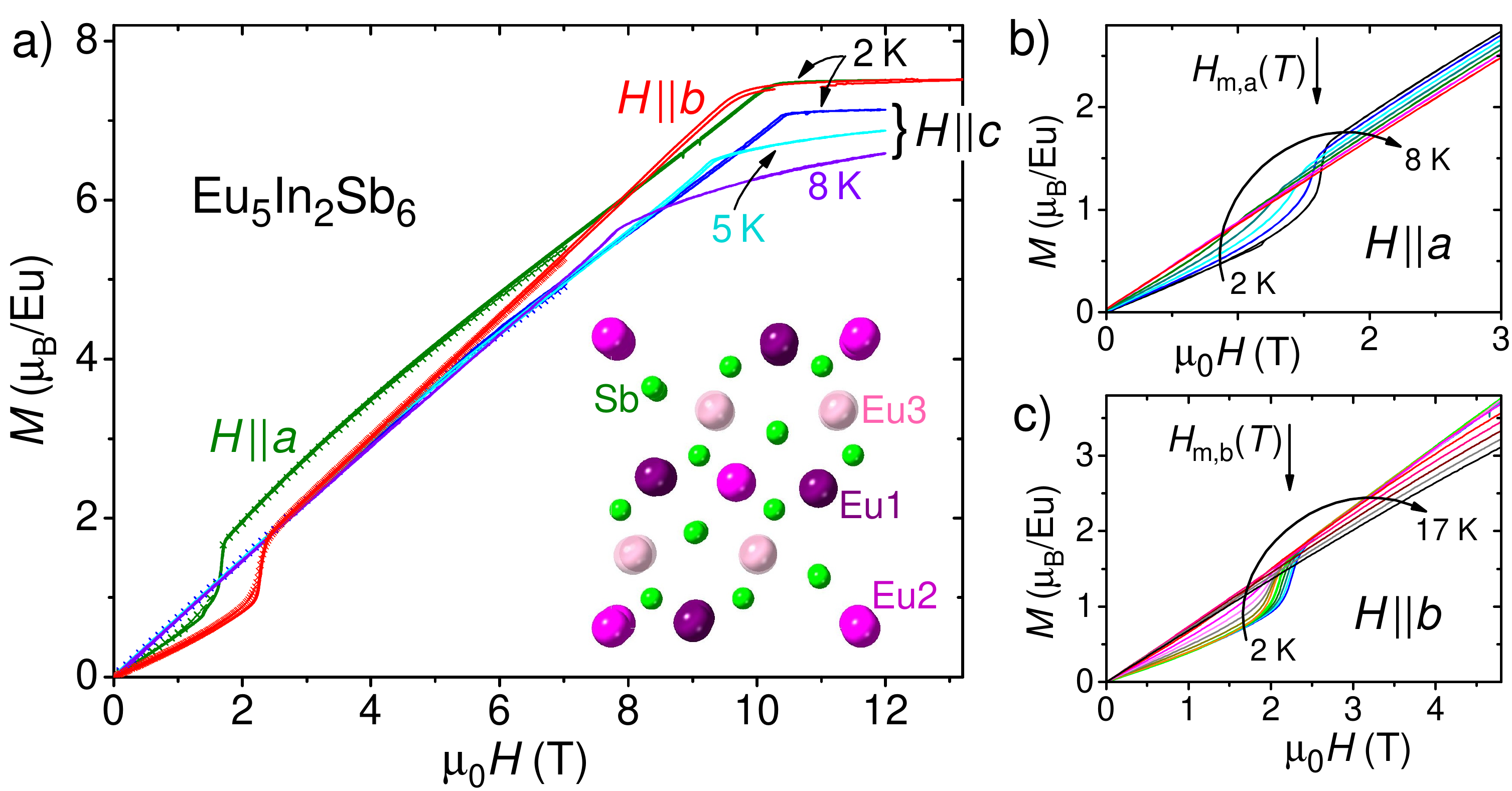}
\caption{(\textbf{a}) Magnetization in dependence on magnetic field for up and
down sweeps. Data at 2 K are shown for fields along all three crystallographic
directions, for $H \parallel c$ also data at 5 K and 8 K are presented. Lines
indicate measurements up to 14 T, markers up to 7 T. Inset: Three groups of
crystallographically different Eu positions (reddish color, marked Eu1, Eu2 and
Eu3) within the crystal structure of {\eis} (Sb in green, In not shown) viewed
along the $c$-axis. (\textbf{b}), (\textbf{c}) Temperature evolution (in steps
of 1~K) of magnetization curves for $H \parallel a$ and $H \parallel b$
emphasizing the metamagnetic transitions at $H_{\rm m}(T)$ (arrows).}
\label{magn}  \end{figure*}
likely also of anisotropic nature. Such electronically inhomogeneous properties
in an intermetallic compound with precise electron count, i.e. in an insulating
environment, provide a rich playground for possibly new quantum states,
specifically when time-reversal symmetry is broken by incorporating magnetic
(here rare earth) elements \cite{cha20} and materials of nonsymmorphic symmetry
are investigated \cite{par13}.

\section*{Results and Discussions}
\subsection*{Magnetic properties}
\begin{figure*}[t]
\centering
\includegraphics[width=0.96\textwidth]{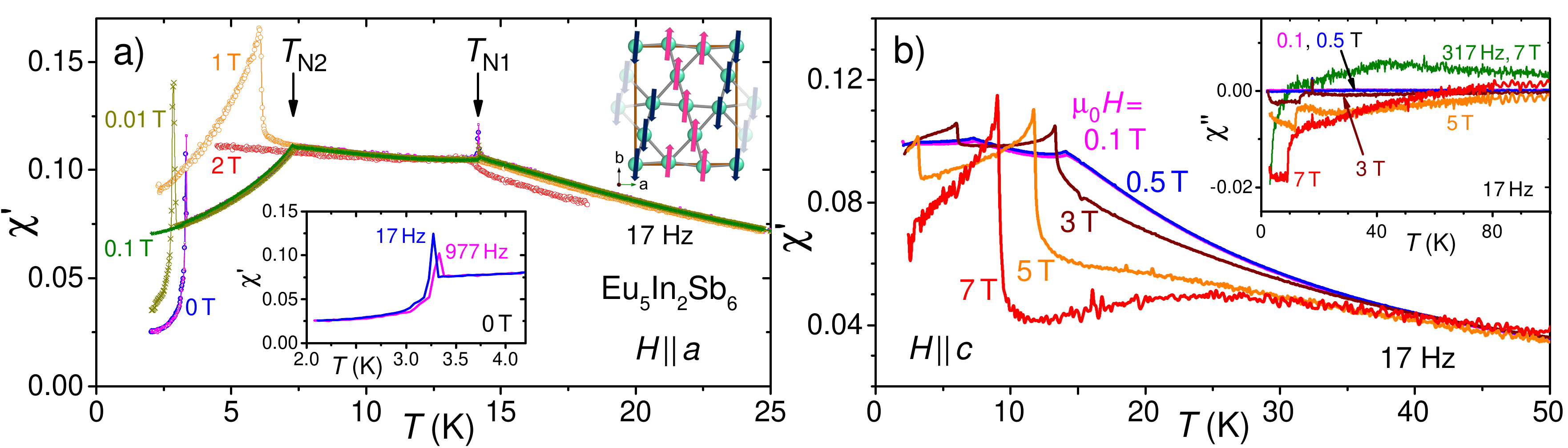}
\caption{(\textbf{a}) Temperature dependent susceptibility $\chi' (T)$ for
different magnetic fields $H$ applied along the crystallographic $a$-axis
illustrating the temperature shifts of the transitions with $H$. Lower inset:
Slow $T$-sweep (0.1 K/min) reveals a small offset for different ac-field (2 Oe)
drive frequencies for the low-$T$ transition. Upper inset: suggested spin
configuration. (\textbf{b}) $\chi' (T)$ for different $H$ along the $c$-axis.
Inset: Imaginary component of susceptibility, $\chi''(T)$ for the same
conditions. For comparison, $\chi''(T)$ at 317 Hz and 7 T is shown. Different
samples were measured in (a) and (b).} \label{suscep}
\end{figure*}

As already mentioned, the crystallographic structure of \eis\ gives rise to
three different Eu sites, cf.\ inset to Fig.\ \ref{magn}(a): Eu(2) and Eu(3)
are surrounded by six Sb (with the octahedra around Eu(3) slightly larger than
for Eu(2)), while Eu(1) has two In and seven Sb close neighbours \cite{par02}.
The nearest Eu distances are between Eu(2)--Eu(3) (3.7987 {\AA}) whereas the
nearest Eu to an Eu(1) is spaced 4.072 \AA\ apart. Magnetic ac susceptibility
$\chi' (T)$ clearly shows two magnetic transitions at $T_{\rm N1} \approx$
14.1~K and $T_{\rm N2} \approx$ 7.2~K \cite{sub16,ros20}, see Fig.\
\ref{suscep}(a) for $H \parallel a$, which are both suppressed in applied
magnetic fields of $\mu_0 H =$ 2\,T.

The magnetization curves $M(H)$ for fields applied along the main
crystallographic axes of \eis\ largely conform to the behavior expected for an
antiferromagnet, Fig.\ \ref{magn}; there is an almost linear increase of
$M(H)$, specifically for $H \parallel c$, and a saturation at high fields.
The saturation value of $M_{\rm sat} \approx 7.0\;\mu_{\rm B}$/Eu for $H 
\parallel c$ is consistent with the expected saturation magnetic moment of
Eu$^{2+}$, $g\mu_{\rm B}J = 7 \mu_{\rm B}$. For $H \parallel a$ and $H 
\parallel b$ slightly higher values of $M_{\rm sat} 
\approx 7.5\;\mu_{\rm B}$/Eu were observed.

It is important to emphasize again the Eu$^{2+}$ state with $L = 0$. In this
case, magnetic anisotropy can be attributed to anisotropic exchange, dipolar
interactions or even crystalline electric fields of higher order. Hence, the
clear anisotropic behavior below $T_{\rm N1}$, as seen in the magnetization
curves $M(H)$, Fig.\ \ref{magn}, for fields applied along the main
crystallographic axes, is likely due to anisotropic exchange between the
Eu-sites. Notably, there are magnetic transitions for $H \parallel a, b$, but
not for $H \parallel c$, which indicates that the magnetic moments are
(primarily) aligned within the $a$-$b$ plane \cite{ale22}. Figures
\ref{magn}(b) and (c) exhibit the disappearance of the metamagnetic transitions
with increasing temperature for $H \parallel a$ and $H \parallel b$,
respectively. Interestingly, this transition is completely vanished at $T
\gtrsim T_{\rm N2}$ for $H \parallel a$, but can be followed up to $T \approx
T_{\rm N1}$ for $H \parallel b$. In the latter case, there is only a weak
shift of the transition field $H_{\rm m}$ up to $T_{\rm N1}$. Furthermore,
there is no notable remanent magnetization found after applying a magnetic
field along any crystallographic direction, as expected for an
antiferromagnetic spin configuration. However, it is essential to note that
the precise magnetic structure of \eis\ has not been reported yet. In the
absence of detailed experimental insight, we consider a predominantly A-type
antiferromagnetic order with antiparallel stacking along $c$, as suggested by
density functional theory (DFT) calculations \cite{ale22}, see upper inset to
Fig.\ \ref{suscep}(a). In all likelihood, the actual spin configuration is
more complex and non-collinear, and may even support very weak ferromagnetism
by spin-canting resulting from Dzyaloshinskii-Moriya interactions allowed in
this low crystallographic symmetry \cite{ros20,ale22,bog02}.

At even lower temperature, $T \lesssim 3$~K, there is a third transition
observed only for $H \parallel a$. This transition is readily suppressed to
below 2 K at fields as small as 0.1~T, and exhibits a very weak frequency
dependence, see lower inset to Fig.\ \ref{suscep}(a). At present, we cannot
rule out In inclusions (resulting from the sample flux growth) causing this
transition.

In line with the above-suggested spin configuration, $\chi' (T)$ exhibits a
different behavior for $H \parallel c$, Fig.\ \ref{suscep}(b): Here, the
transition at $T_{\rm N1}$ becomes ever more pronounced while shifting to lower
temperature with increasing dc-field $H$. Also, the transition at $T_{\rm N2}$,
albeit shifted down to about 3.3~K, can still be recognized at 5~T. The shallow
maximum of $\chi'(T \sim \rm{25\; K}, \mu_0 H = \rm{7\; T})$ and the peculiar
$T$-dependence of the imaginary component of the susceptibility, $\chi''(T)$,
inset of Fig.\ \ref{suscep}(b), will be discussed below.

\begin{figure}[b]
\centering
\includegraphics[width=0.44\textwidth]{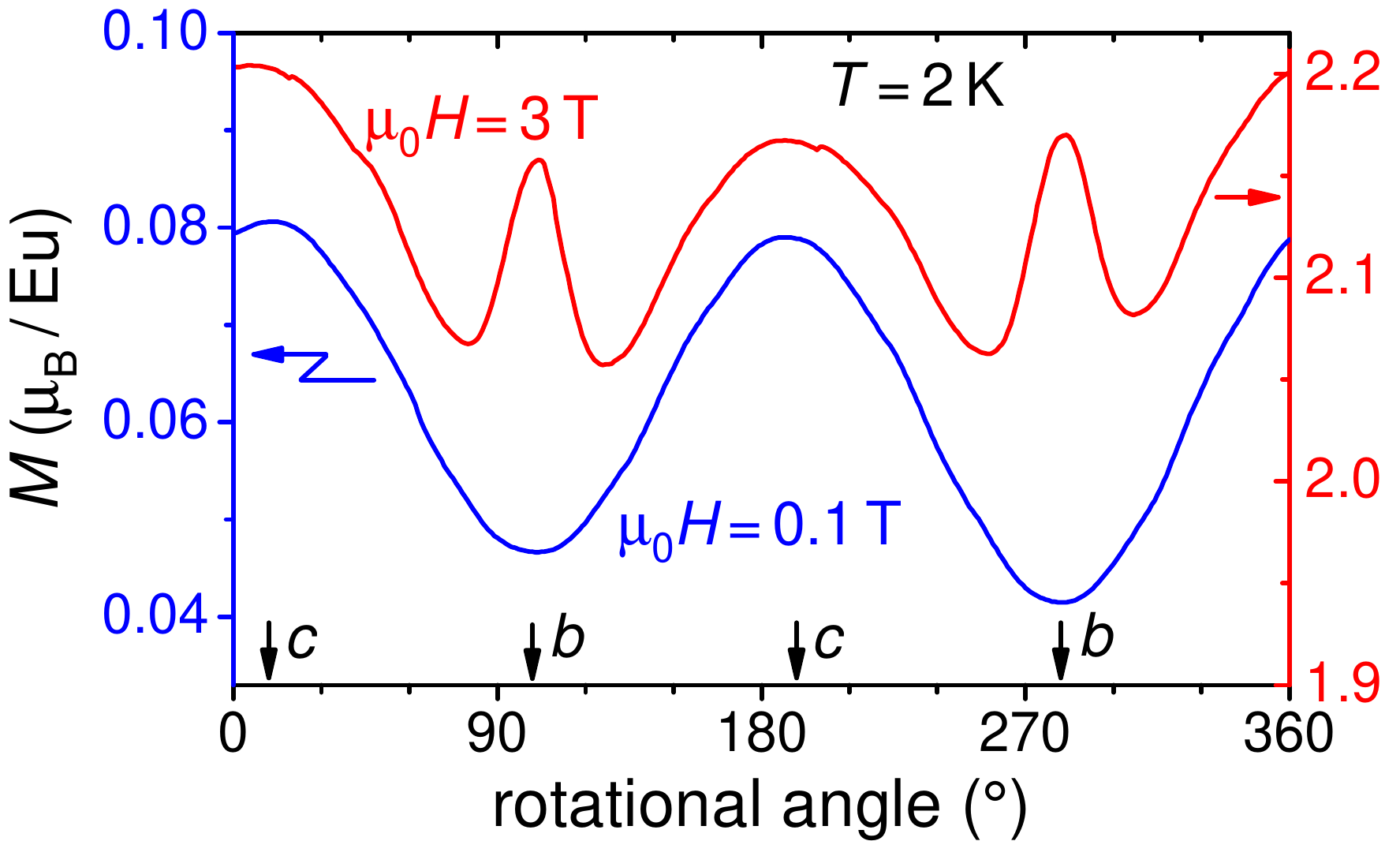}
\caption{Magnetization in dependence on sample orientation with respect to
$H$ for two applied fields ($\mu_0 H =$ 0.1 T, 3 T) below and above
$\mu_0 H_{\rm m} \approx 2.2$ T at $T = 2$ K. The sample is rotated in the
$b$--$c$ plane, i.e.\ around the $a$-axis, with the alignment of the
respective axis with the direction of $H$ marked by arrows.} \label{Mrot}
\end{figure}

The numerous transitions and different dependencies of $\chi'$ certainly
testify to an intricate response of \eis\ to magnetic fields. To gain further
insight, the isothermal magnetization was measured upon rotating the sample in
constant applied fields. The results of one exemplary measurement is presented
in Fig.\ \ref{Mrot} for fields of $\mu_0 H =$ 0.1, 3~T and at $T = 2$~K, with
the sample being rotated around its crystallographic $a$ axis (for more results
on sample rotation see Supplementary Information, Fig.\ S1). At small fields
(0.1~T), we find the expected two-fold symmetry: There is little response of
the local moments for $H \parallel b$, where $b$ is likely the easy
magnetization direction \cite{ros20,ale22} along which the moments are already
aligned in the ground state, while the moments can more easily be turned
towards the $c$ direction even by a small field. At $\mu_0 H =$ 3~T, the curve
is just offset for a broad angular range around the $c$ direction. In addition,
however, there is a pronounced maximum for $H \parallel b$, which is
quickly suppressed within a range of $\pm 20^{\circ}$. This observation, along
with the jump in magnetization at $H_{m}$ for $H \parallel b$ shown in Fig.\
\ref{magn}(c), is a clear indication for a field-induced change in the magnetic
structure (possibly a spin-flop transition). This assignment is further
supported by the very weak temperature dependence of $H_{m}(T)$ and is
consistent with the $b$ axis being the magnetically easy direction.

There are several indications for \eis\ being a candidate material for the
formation of magnetic polarons near the magnetic ordering temperature, ranging
from colossal magnetoresistance \cite{ros20} and piezoresistance \cite{gho22}
to characteristic changes of the Eu$^{2+}$ electron spin resonance (ESR)
\cite{sou22}. Here, magnetic polarons refer to magnetically ordered clusters
within which conduction electrons are localized via strong exchange interaction
with the 4$f$ moments, giving rise to spin-polarization \cite{mol07}. The
$\chi' (T)$ data for small applied field exhibit a tiny deviation from
\begin{figure*}[t]
\centering
\includegraphics[width=0.8\textwidth]{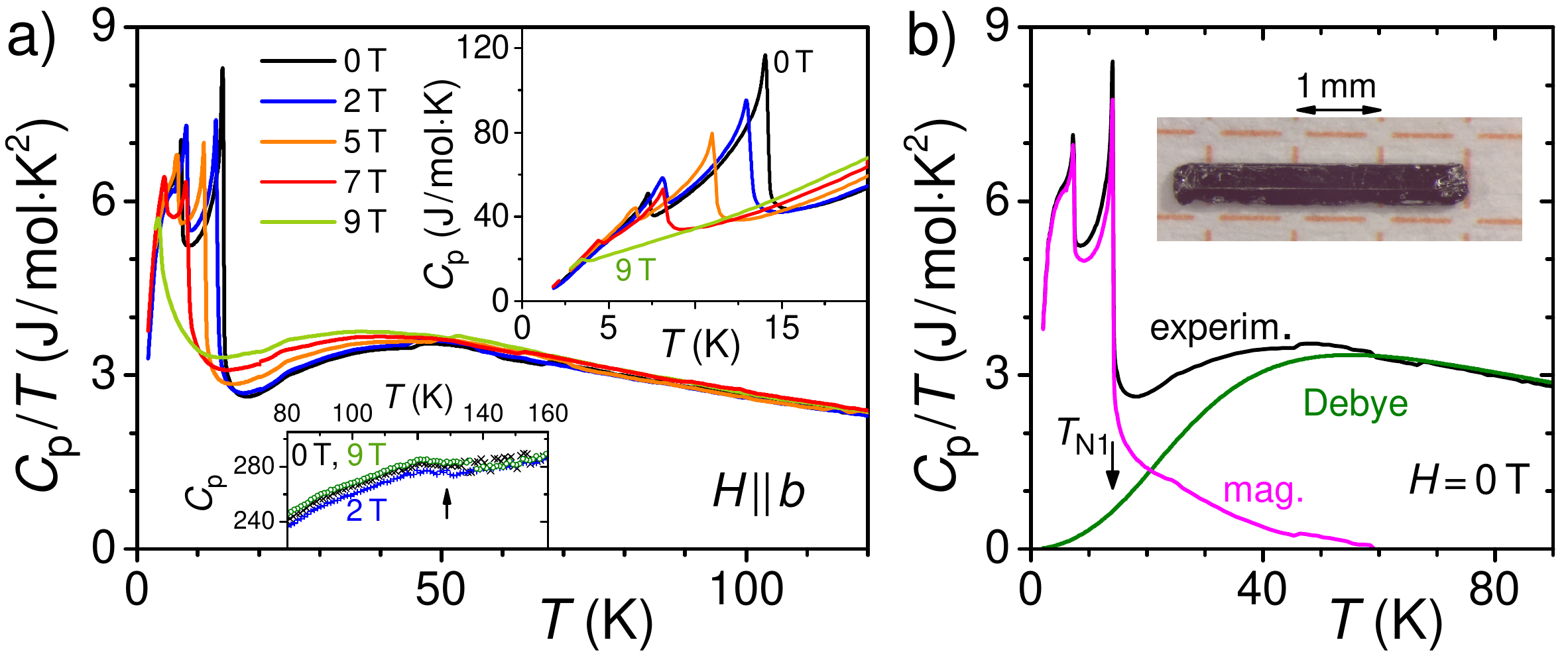}
\caption{(\textbf{a}) Specific heat divided by temperature, $C_p / T$, in
dependence on $T$ for applied fields $H \parallel b$ up to 9~T. The insets show
enlarged views of $C_p(T)$ within certain temperature ranges (in units of
J/mol\,K). (\textbf{b)} $C_{\rm ph}$ contribution from a Debye model (green
curve) fitted to the experimental $C_p(T)$-data (black) within 60~K $\leq T
\leq$ 200~K and used to estimate $C_{\rm mag} / T$ (magenta). Inset: photograph
of one representative \eis\ sample with the long dimension corresponding to the
$c$ axis.} \label{specheat}
\end{figure*}
Curie-Weiss behavior below about 180 K, and a more obvious one around
50--60 K (see Supplementary Information, Fig.\ S2) which was argued to
be caused by the formation of magnetic polarons and correlations between them,
respectively. Very likely, the polarons grow in size with increasing applied
field such that correlations become more pronounced. Consequently, the
deviations from Curie-Weiss behavior are stronger for larger field, even
resulting in a shallow maximum of $\chi'(T)$ at 7~T, see Fig.\ \ref{suscep}(b).
This picture is corroborated by the imaginary component of the susceptibility,
$\chi''(T)$, presented in the inset of Fig.\ \ref{suscep}(b). $\chi''$ is
related to dissipative magnetic processes and is typically zero in
antiferromagnetic phases. Interestingly, for large enough applied fields,
$\chi''(T \lesssim 50 {\rm K})$ is finite, indicating ferro- or ferrimagnetic
contributions, and negative, consistent with an inhomogeneous state of
magnetization \cite{uly20}. Moreover, $\chi''$ is positive, if measured at
317~Hz and 7~T, emphasizing a changed response to higher drive frequencies.

The behavior just described is also observed when the magnetic field is applied
along the crystallographic $b$ axis. In comparison to other compounds for which
a magnetic polaron scenario is considered, the phenomenology found in \eis\ is
consistent with findings for EuB$_6$ (see, e.g., susceptibility data in Ref.\
\cite{zha09}), but differs from the one reported in Sr doped LaCoO$_3$ samples
\cite{kum20}.  In the latter case, the spin-glass behavior of ferromagnetic
clusters embedded in a non-ferromagnetic matrix is thought to be revealed by a
frequency-dependent cusp in the ac susceptibility data.

\subsection*{Heat capacity}
Results of the measurements of the specific heat $C_p (T,H)$ for different
magnetic fields $H \parallel b$ are presented in Fig.\ \ref{specheat}. Here,
the main panel shows $C_p / T$ as a function of $T$, while the insets zoom into
$C_p$-data within two temperature ranges of interest. The two transitions at
$T_{\rm N1}$ and $T_{\rm N2}$ as well as their suppression in magnetic field
are clearly revealed. Importantly, the impact of applied magnetic fields can be
recognized up to temperatures of about 60~K whereas at higher temperatures
there are no obvious differences between the data obtained at different
magnetic fields. The experimentally observed zero-field $C_p / T$-data exhibit
a pronounced maximum at $T_{\rm max} \sim 45$\,K which shifts to lower
temperatures with increasing magnetic field ($T_{\rm max}^{\rm 9 T}
\sim 37$\,K). Consequently, we will relate this maximum to the magnetic
properties of \eis.

To analyse $C_p$, magnetic, electronic and phonon contributions to the specific
heat are considered, $C_p(T,H) = C_{\rm mag} + C_{\rm el} + C_{\rm ph}$. Here,
$C_{\rm el} = \gamma\,T$ and $C_{\rm ph}$ are described by a linear-in-$T$ and
a Debye model, respectively. Given the number of parameters, the above
description of $C_p$ cannot unambiguously be fitted to our experimental data.
Therefore, $(C_{\rm el} + C_{\rm ph})$ was fitted to the experimental data only
for 60~K$\,\leq T \leq\,$200~K, i.e.\ in a $T$-range above the temperature
below which differences for different magnetic fields were seen. The fit yields
a Debye temperature of $\theta_{\rm D}=$ 197~K [green curve in Fig.\
\ref{specheat}(b)] and a vanishing value for $\gamma$. Using these fit results
to subtract the corresponding contributions from the experimental data in the
low-temperature range, $C_{\rm mag} / T = C_p/T - C_{\rm ph}/T$ shown as
magenta curve in Fig.\ \ref{specheat}(b), reveals a significant magnetic
contribution to the specific heat even above $T_{\rm N1}$. A very similar
behavior was observed for a second set of samples measured up to 280~K to
improve on the Debye fit, see Supplementary Information Fig.\ S3. Integrating
$C_{\rm mag}/T$ up to $T = 15$~K, i.e. just above $T_{\rm N1}$ but well below
$T_{\rm max}$ yields about 80\% of the expected magnetic entropy $S_{\rm mag}
= 5R\ln (2S+1) \approx$ 86.4 J/mol\,K (where $R$ and $S$ are the gas constant
and the spin momentum of Eu$^{2+}$, respectively), i.e. the integration has to
be carried out up to higher temperatures to recover the full magnetic entropy.
This inference holds despite the uncertainty of our analysis as well as for
\begin{figure*}[t]
\centering
\includegraphics[width=0.94\textwidth]{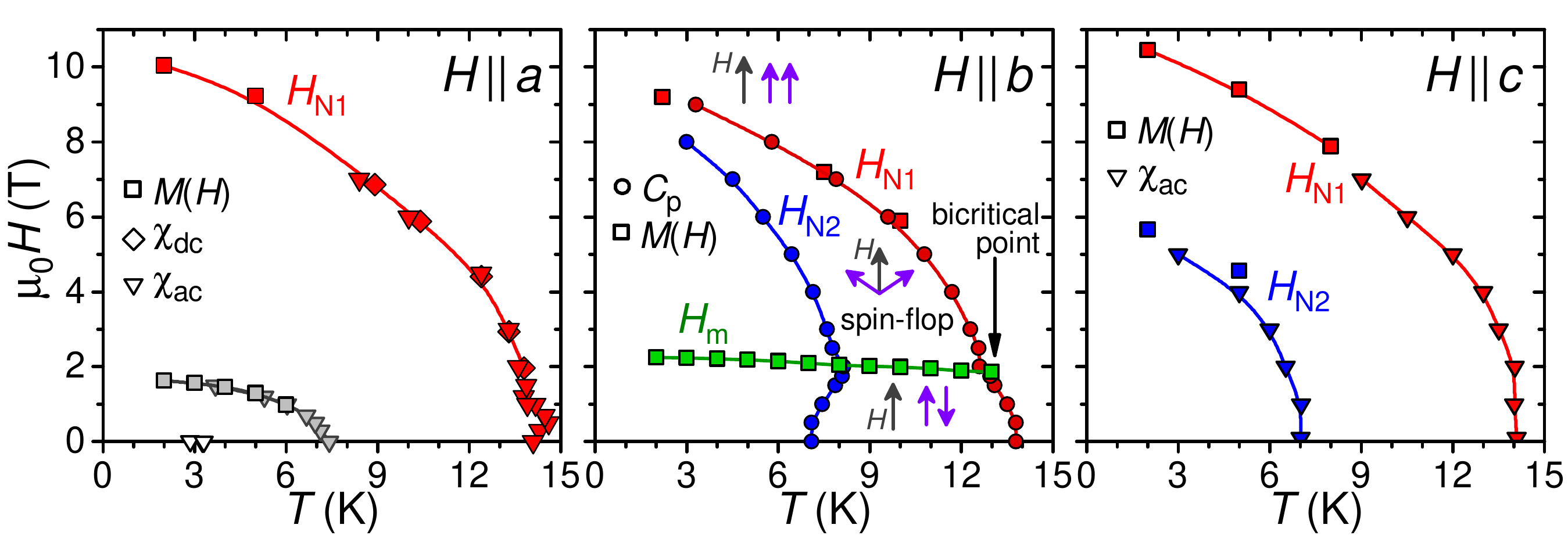}
\caption{$H$--$T$ phase diagrams for magnetic fields aligned along different
crystallographic directions, \textbf{left}: $H \parallel a$, \textbf{center}:
$H \parallel b$, \textbf{right}: $H \parallel c$, as extracted from ac and dc
susceptibility $\chi$, magnetization $M$ and specific heat $C_p$ measurements.
Violet arrows depict a possible magnetization orientation with respect to
magnetic field (gray arrows). To enable comparison, sample demagnetization
effects are taken into consideration.} \label{phasedia}
\end{figure*}
our data in magnetic field (see Supplementary Information Fig.\ S3) and
reinforces our conclusion above that the feature related to $T_{\rm max}$ is
magnetic in nature.

Near $T_{\rm max} \sim 45$\,K a strong deviation of the susceptibility from a
Curie-Weiss law was observed (Fig.\ S2) which was previously ascribed to the
onset of interactions between polarons \cite{ros20}. Following this
interpretation, we assume this maximum to result from a \emph{field-dependent} Schottky-like anomaly. Indeed, there are examples of such Schottky-like
anomalies, where the temperature $T_{\rm max}$ decreases with increasing
magnetic field \cite{adh19}. Also, a low-temperature Schottky anomaly in
LaCoO$_3$ has been previously linked to magnetic polarons \cite{he09}. If an
interpretation of $T_{\rm max}$ based on a polaron-interaction scenario is
correct, the \emph{formation} of polarons is expected to take place at
significantly higher temperatures. Here we note that there is a small hump
observed in $C_p(T)$ at $T_{\rm hump} \sim 130$\,K, lower inset to Fig.\
\ref{specheat}(a), a temperature which is somewhat lower compared to the one
expected for polaron formation from susceptibility, Fig.\ S2, and Refs.\
\cite{ros20,sou22}.

A Schottky contribution $C_{\rm Sch}$ to $C_p$ is associated with a two level
system \cite{gop66}. We speculate that such a two-level scenario may result
from ferromagnetic (fm) correlations related to the magnetic polarons on the
one hand, and antiferromagnetic interactions between Eu ions on the other hand.
The latter are expected to prevail at lower temperature near $T_{\rm N1}$ while
fm correlations are likely the dominant ones at higher temperatures, where they
are expected to be stabilized in magnetic fields. Indeed, assuming a lattice
background independent of magnetic field and fitting the specific heat data
well above $T_{\rm N1}$ with a Debye term and a magnetic field-dependent
Schottky contribution reveals a systematic increase of the level splitting from
$\delta \sim 2.2$\,meV at $\mu_0 H = 0$ to about 4.5\,meV at 9\,T, while at the
same time the magnitude of the magnetic Schottky contribution shows a
$\sim 30$\,\% decrease with increasing field. This explains the apparent shift
of $T_{\rm max}$ in $C_p/T(T)$ to lower temperatures with increasing field.

Albeit a Schottky anomaly linked to polaron interaction is consistent with
the observed temperature evolution of $C_p$, we cannot exclude alternative
explanations. For instance, a temperature well above $T_{\rm N1}$ required to
gain the full magnetic entropy might also result from some partial ordering
due to frustration between different Eu sites. We wish to stress, however,
that---independent of any fitting or interpretation---there is a considerable
amount of magnetic contribution to $C_p$ found at temperatures well above
$T_{\rm N1}$.

\subsection*{Magnetic phase diagrams}
In order to gain more insight into the magnetic behavior we construct $H$--$T$
phase diagrams for the different crystallographic directions making use of our
magnetization, susceptibility and specific heat data. The resulting phase
diagrams for $H \parallel a$ and $H \parallel c$ are qualitatively similar,
with $c$ clearly being the magnetically hard direction. We note that for
$H \parallel a$ the jump of $M(H)$ at 2~K and $\sim$1.6~T [Fig.\ \ref{magn}(b)]
is indicative of a metamagnetic transition, similar to the observation for
$H \parallel b$ [Fig.\ \ref{magn}(c)]. However, in contrast to the latter case,
the transition for $H \parallel a$ is readily suppressed near $T_{\rm N2}$.
This makes an assignment of this transition for $H \parallel a$ to $H_{\rm m}$
or $H_{\rm N2}$ difficult. Moreover, only in case of $H\parallel a$ is there a
slight change of slope of $M(H)$ near 5~T, Fig.\ \ref{magn}(a), and a third
transition seen in $\chi'(T)$ at $\mu_0 H \leq 0.1$~T, Fig.\ \ref{suscep}(a).

A more complex behavior is found for the $b$ axis, i.e.\ the magnetically easy
direction, where a field-induced change in the magnetic structure (likely a
spin-flop transition) at $H_{\rm m}$ is observed. This gives rise to a
bicritical point, where $H_{\rm m}(T)$ terminates at $H_{\rm N1}(T)$. Here,
$T_{\rm N1}(H)$ exhibits the characteristic kink at the bicritical point
\cite{fis74}. The non-monotonic $H$-dependence of $T_{\rm N2}(H)$ indicates a
\begin{figure*}[t]
\centering
\includegraphics[width=0.99\textwidth]{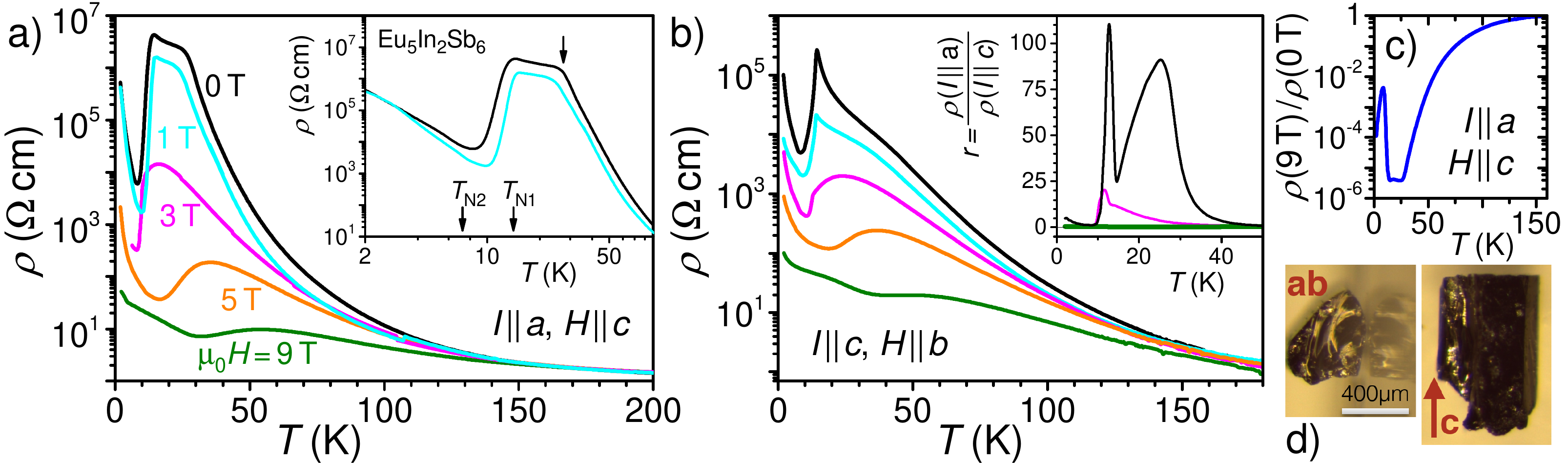}
\caption{Resistivity $\rho(T)$ in dependence on temperature for current $I$
along different directions: \textbf{(a)} $I \parallel a$ and $H \parallel c$
and \textbf{(b)} $I \parallel c$ and $H \parallel b$. Inset in \textbf{(a)}:
Zoom into the low-$T$ range on a logarithmic $T$-scale. At zero field, a kink
is observed around 27~K (arrow). Inset in \textbf{(b)}: Ratio $r$ of
resistivities for the different configurations shown in (a) and (b). The
zero-field maximum near 27~K is suppressed with field. At $H = 9$~T, $r$ is
of order 0.3 (green curve). All panels share the same color code for magnetic
fields. \textbf{(c)} Ratio of resistivities measured at fields of 9~T and 0~T,
as shown in (a). \textbf{(d)} Photographs of the sample, scale bar 400~$\mu$m.}
\label{resist} \end{figure*}
stiffening of the ordering taking place at $T_{\rm N2}$ for fields below the
spin-flop transition, while above $H_{\rm m}$ the ordering is successively
suppressed with increasing field. One may speculate that a similar interplay of
exchange interactions is at play for $T_{\rm N1}(H)$ at very small fields
$H \parallel a$. All these observations point to a complex magnetic exchange
and ordering mechanism in \eis\ as can be expected from the crystallographic
structure.

\subsection*{Resistivity measurements}
Electronic transport in general, but specifically in a material as complex as
\eis\, can be influenced by a number of factors (this may also include In
inclusions originating from sample growth). Indeed, resistivity measurements
carried out on different samples exhibited somewhat different
$\rho(T)$-results, in particular for $T < T_{\rm N1}$ (compare results shown
in Fig.\ \ref{resist} and Supplementary Information Fig.\ S4). Therefore,
resistivity data were not included in constructing the phase diagrams of Fig.\
\ref{phasedia}. Rather, we make use of the insight gained from other
measurements so far when discussing the electronic transport results in the
following.

In Fig.\ \ref{resist} results of the transversal resistivity $\rho(T,H)$ are
compared for currents $I$ along the $a$ and the $c$ direction,
respectively. Clearly, the strong anisotropy of \eis\ is also seen in
$\rho(T)$, specifically in zero field. For, $I \parallel a$ the resistivity
$\rho_{I\parallel a}(T)$ rises much more rapidly upon cooling than
$\rho_{I\parallel c}(T)$. This behavior (also seen in Fig.\ S4) may be
related to the crystal structure of \eis\ which is characterized by inﬁnite
[In$_2$Sb$_6$]$^{10−}$ ribbons oriented parallel to the crystallographic $c$
axis \cite{par02} and hence, favors transport for $\rho_{I \parallel c}(T)$.
Near 27 K, $\rho_{I\parallel a}(T)$ exhibits a kink into a much less-steep
behavior [see arrows in insets to Fig.\ \ref{resist}(a) and Fig.\ S4]
before dropping by several orders of magnitude below $T_{\rm N1}$.
For $I\parallel c$, a smaller hump is detected, in agreement with Ref.\
\cite{ros20}. In consequence, the ratio of resistivities, $r(T) = (\rho_{I
\parallel a})/(\rho_{I\parallel c})$ measured for the two different current
directions exhibits a pronounced maximum near 27 K, inset of Fig.\
\ref{resist}(b). The second, sharper maximum at 12.7~K is likely related to
the antiferromagnetic order at $T_{\rm N1}$. Larger spin fluctuations can be
expected in the plane perpendicular to the magnetically hard $c$ direction,
increasing $\rho_{I \parallel a}$, an effect that quickly subsides for
decreasing $T$, i.e.\ away from the magnetic transition.

As mentioned above, $\chi'(T)$ exhibits a deviation from Curie-Weiss behavior
below about 50~K (Fig.\ S2). Therefore, the rise of $r(T)$ for $T \lesssim
50$~K is very likely linked to the magnetic behavior of \eis, in line
with the specific heat results discussed above. More specifically, ellipsoidal
polarons have been suggested for \eis\ which start to interact below about 50~K
\cite{gho22}. We speculate that these ellipsoidal polarons are more extended
perpendicular to the magnetically hard $c$ direction and strongly interact
below $\sim$27~K, possibly causing the plateau-like behavior of $\rho_{I
\parallel a}(T)$. A further growth of the polarons upon decreasing $T$ also
increases interaction along the $c$ direction resulting in the drop of $r(T)$
below about 25~K. Applying magnetic fields is expected to level out
inhomogeneities in the magnetic state of \eis. Our finding of a significantly
suppressed $r(T)$ already at 3~T and an almost flat behavior at 9~T, inset of
Fig.\ \ref{resist}(b), strongly supports such a polaron scenario.

As expected in a polaron scenario, large MR effects are observed \cite{ros20},
especially in a temperature range $T_{\rm N1} \lesssim T \lesssim 27$~K. This
can be inferred from the ratio of resistivities $\rho (9\, {\rm T}) / \rho
(0\, {\rm T})$, plotted in dependence on $T$ for $I \parallel a$ in Fig.\
\ref{resist}(c). This ratio reaches values well below 10$^{-5}$, corresponding
to a CMR $[$defined as $(\rho(H) - \rho(0)) / \rho(0)]$ of more than
$-$99.999\%. The anisotropy of $\rho$ in \eis\ is also reflected in the MR:
For $I \parallel c$, the ratio $\rho (9 {\rm T}) / \rho (0 {\rm T})$ goes only
down to $\sim\! 1.6 \cdot 10^{-4}$ close to $T_{\rm N1}$.

One characteristic of materials exhibiting polaron formation is a low carrier
concentration $n$ such that energy gain by carrier localization is achieved
\cite{mol07}. Here, Hall measurements are called for. However, the
determination of $n$ is straightforward only if inhomogeneous magnetic states
and spurious contributions to the Hall signal are avoided. Within this
limitation (i.e.\ for $\mu_0 H \geq$ 7~T where a linear Hall response was
observed) we obtain $n \approx 1.5 \cdot 10^{17}$ cm$^{-3}$ at $T = 100$~K
assuming a single (hole-like) band. Note that this corresponds to about 0.02
carriers per Eu site, a number very similar \cite{gru85} to the one found for
EuB$_6$ which is a prototypical material for polaron formation
\cite{nyh97,poh18}.

\section*{Conclusions}
The CMR effect observed in \eis\ is very likely related to the formation of
magnetic polarons, similar to EuB$_6$ or the doped manganites. However, there
are also marked differences: i) As seen from Fig.\ \ref{resist}(c) the MR in
\eis\ is always negative whereas the MR in EuB$_6$ and doped manganites turns
positive below the magnetic ordering temperature (for discussions, see Refs.\
\cite{cal04,coe99}). Likely, this is related to the ferromagnetic ordering in
the latter two materials while \eis\ orders antiferromagnetically. Therefore,
\eis\ provides a rare opportunity to study magnetic polaron formation in an
antiferromagnetic environment. ii) The large MR effects in \eis\ extend to
comparatively high temperatures. The resistivity is still suppressed by an
\emph{order of magnitude} in $\mu_0 H = 9$~T at around 75~K, i.e.\ at more than
six times $T_{\rm N1}$. By comparison, the polaron formation in EuB$_6$ only
\emph{sets in} at around 35~K \cite{nyh97,man14}, which corresponds to about
three times the ferromagnetic ordering temperature. iii) The highly anisotropic
magnetic and transport properties of \eis\ are certainly related to its complex
crystallographic structure. This is in line with the observation of highly
anisotropic  electronic transport properties in the manganites with layered
perovskite structure \cite{mor96}.

The latter, anisotropic properties stem from anisotropic exchange interactions,
given the magnetically isotropic Eu$^{2+}$ ground state configuration
$^8 S_{7/2}$. Earlier spin-resolved band structure calculations suggested a
spin conﬁguration with ferromagnetically coupled Eu moments in the $ab$ plane,
which are alternatingly stacked in an antiparallel staggered order
\cite{ale22}. Our angular dependent magnetization measurements support such a
picture. In particular, a spin-flop transition for $H \parallel b$ indicates a
preferred alignment of magnetic momentums along the $b$ direction.

The here established phase diagrams can guide future exploration of this
complex material. Clearly, the magnetic structure needs to be resolved further,
including a possible weakly ferromagnetic component resulting from spin canting
due to Dzyaloshinskii-Moriya exchange. The insights provided here are essential
for an evaluation of the surface topology in this nonsymmorphic material. In
particular, \eis\ may provide a test case for studying the impact of electronic
inhomogeneities on topology \cite{zha18,pan21}.

\section*{Methods}
Single crystalline samples \eis\ were grown by a combined In-Sb self-flux
technique \cite{ros20}. The crystallographic structure was confirmed by x-ray
diffractometry. For sample orientation a real-time Laue x-ray system was
employed (Laue-Camera GmbH). For magnetization measurements up to magnetic
fields of 14 T a vibrating sample magnetometer (VSM, Oxford Instruments Inc.)
was utilized. Magnetic measurements up to 7~T were also conducted using a
magnetic property measurement systems (MPMS3, Quantum Design Inc.). If not
noted otherwise, susceptibility was measured with an applied ac field of
10 Oe after cooling in zero applied field (ZFC). The angle dependence of
magnetization was measured by means of a horizontal rotator inside an MPMS-XL,
Quantum Design Inc. Specific heat was measured using a physical property
measurement system (PPMS, Quantum Design Inc.) equipped with a calorimeter
that utilizes a quasi-adiabatic thermal relaxation technique. Electronic
transport was investigated using the same PPMS system. In an effort to
accurately measure the high sample resistances at low temperatures, an
external, lock-in-based circuitry hooked up to the PPMS was implemented. In
addition, to enable electronic transport measurements, a somewhat thicker
sample was used, Fig.\ \ref{resist}(c).

\section*{Data availability}
Full sets of crystallographic data generated and/or analysed during the
current study have been deposited with the joint CCDC and FIZ Karlsruhe
deposition service. The data can be obtained free of charge from The Cambridge
Crystallographic Data Centre via www.ccdc.cam.ac.uk/structures citing
deposition number CSD 2214610. All other data generated and analysed during
this study are included in this published article (and its supplementary
information files) and/or are available from the corresponding author on
reasonable request.

\section*{Acknowledgements}
The authors thank U.\ K.\ R\"{o}{\ss}ler for insightful discussions, Tim Thyzel
for help with the analysis of the specific heat data, Ralf Koban for technical
assistance and Horst Borrmann for help with the data deposition. SR
acknowledges support by the Deutsche Forschungsgemeinschaft (DFG, German
Research Foundation) through SFB 1143. Work at the Max-Planck-Institute for
Chemical Physics of Solids in Dresden and at Goethe University Frankfurt was
funded by the Deutsche Forschungsgemeinschaft (DFG, German Research
Foundation), Project No. 449866704. Work at Los Alamos was performed under the
auspices of the U.S.\ Department of Energy, Office of Basic Energy Sciences,
Division of Materials Science and Engineering. MSC acknowledges support from
the Laboratory Directed Research and Development program.

\section*{Author contributions}
S.R., P.F.S.R. and S.W. conceived the experiments, P.F.S.R. prepared the
samples, M.V.A.C., S.R., S.G., M.D. and M.S.C. conducted the experiments,
M.V.A.C., S.R., J.M. and S.W. analysed the results, M.V.A.C., J.M. and S.W.
wrote the manuscript. All authors reviewed the manuscript.

\section*{Additional Information}
\textbf{Supplementary information} accompanies this paper at
http://www.nature.com/srep\\

\noindent
\textbf{Competing interests:} The authors declare no potential conflict of
interest.
\end{document}